\documentclass[aps,prd,twocolumn,showpacs,groupedaddress,nofootinbib,longbibliography]
{revtex4-1}

\expandafter\let\csname equation*\endcsname\relax 
\expandafter\let\csname endequation*\endcsname\relax 

\usepackage{xspace}
\usepackage{stmaryrd}
\usepackage{times}
\usepackage{amsfonts}
\usepackage{graphicx}
\usepackage{amsmath}
\usepackage{amssymb}
\usepackage{color}
\usepackage{subfigure}
\usepackage{natbib}
\usepackage{tensor}
\usepackage{accents}
\usepackage[colorlinks=true, citecolor=blue]{hyperref}
\usepackage{url}

\def\beq{\begin{equation}}
\def\eeq{\end{equation}}
\def\bea{\begin{eqnarray}}
\def\eea{\end{eqnarray}}
\def\ben{\begin{enumerate}}
\def\een{\end{enumerate}}
\def\la{\langle}
\def\ra{\rangle}

\def\d{\delta}

\def\o{\omega}

\def\L{\Lambda}

\def\half{{\textstyle{\frac{1}{2}}}}

\def\cA{${\cal A}_{\partial}$}

\begin{document}

\title{Diffeomorphism invariance and the black hole information paradox}

\author{Ted Jacobson and Phuc Nguyen}
\affiliation{Maryland Center for Fundamental Physics, 
University of Maryland, College Park, MD 20742}

\begin{abstract}

We argue that the resolution to the black hole information paradox
lies in a proper accounting of the implications of diffeomorphism invariance 
for the Hilbert space and observables of quantum gravity. The setting of asymptotically
Anti-de Sitter spacetime is adopted for most of the paper, but in 
the framework of canonical quantum gravity, without invoking AdS/CFT duality. 
We present Marolf's argument that boundary unitarity is a consequence of diffeomorphism invariance, and 
show that its failure to apply in the classical limit results from a lack of analyticity 
that has no quantum counterpart. 
We argue that boundary unitarity leads to a {\it boundary information paradox}, which 
generalizes the black hole information paradox and
arises in virtually any scattering process.
We propose a resolution that involves 
operators of the boundary algebra that redundantly encode information about physics in the bulk,
and explain why such redundancy need not violate the algebraic no cloning theorem.
We also argue that the infaller paradox, which has motivated the firewall hypothesis 
for black hole horizons,  is ill-posed in quantum gravity, because it ignores
essential aspects of the nature of the Hilbert space and observables in quantum gravity.

\end{abstract}

\maketitle

\section{Introduction}

The black hole information paradox 
holds promise for teaching us something fundamental about quantum gravity. 
The paradox concerns how to reconcile well-established principles of local quantum field theory
with expectations from quantum gravity. 
We currently lack  direct experimental evidence of quantum gravity, so the paradox 
provides a welcome challenge to our understanding of the foundations of quantum 
gravity, in particular of the nature of its states and observables. 
To be sure, despite many decades of research and discussion, it is not universally 
agreed that any paradox exists in the first place \cite{Unruh:1995gn, Unruh:2017uaw, Wallace:2017wzs}. 
In this  paper we review compelling arguments 
that it does exist, and we argue that it can be resolved by paying close attention to the role of 
diffeomorphism invariance in the theory. 
Our conclusion is not that local QFT or quantum gravity has some unexpected breakdown, but that 
we just need to better understand the quantum mechanical framework of quantum gravity.

The information paradox has many faces. The one we focus on in this paper is raised by Marolf's argument for {\it boundary unitarity} \cite{Marolf:2008mf}. This argument is founded on the assumption that the diffeomorphism invariance of classical general relativity extends to quantum gravity. The argument is agnostic about the form taken by the UV completion of the theory, as long as it preserves the classical property---which follows from diffeomorphism invariance---that the Hamiltonian is a boundary term. In that case, the algebra of boundary observables includes the Hamiltonian. We consider the case of asymptotically AdS spacetime, so that the boundary is timelike. Then, the
fact that the Hamiltonian is a boundary observable arguably implies that the boundary algebra evolves continuously into itself. If that is so, then information available at the boundary at one time is also available at any other time. 
Note that, although we focus on the example of asymptotic AdS boundary conditions,
we are {\it not} invoking AdS/CFT duality, or any particular UV completion of the the quantum gravity
theory (although it may be that such a duality is inevitably what a UV completion looks like).

One of our aims in this paper is to support this argument that the boundary algebra evolves into itself, and to explain why the argument fails to apply in the classical case. This will hinge on a property of analyticity that we will establish.
Our other aim is to resolve what we will call the ``boundary information paradox" (BIP). This paradox is the apparent contradiction between unitary evolution of boundary observables, and the presence of 
bulk degrees of freedom,
some of which propagate to the boundary while 
being entangled with
others that remain in the bulk. 
A key point we shall argue is that the BIP arises in generic processes, not only ones involving black holes. It is an information paradox, but not a black hole information paradox. We regard the BIP
as a training ground for understanding the Hilbert space and observables of  quantum gravity.

We postpone to the end of the paper a consideration of  
the ``infaller paradox", wherein  local bulk physics apparently exhibits violation of the monogamy of entanglement in the neighborhood of a black hole horizon. Having developed some understanding of the Hilbert space and observables of
quantum gravity, we will argue that the infaller paradox is actually ill-posed, so that it is not clear 
whether there is any such paradox at all. 

We do {\it not} make use of reduced phase space quantization, nor do we make use of AdS/CFT duality, other 
than as an example. Let us explain why not. 
The bulk/boundary dichotomy at the center of our considerations would be obscured if we were to adopt the framework of reduced phase space quantization, where all gauge freedom is eliminated before quantization.
Instead, we adopt the framework of Dirac quantization, in which the physical Hilbert space is the subspace of a kinematic Hilbert space that is annihilated by the gauge constraint operators, and in which physical (i.e.\ guge invariant) observables commute with the constraints. 
In the reduced phase space approach, the question of boundary unitarity just doesn't even come up.
We elect to consider the Dirac quantization,
because it poses puzzles that are hopefully instructive concerning the nature of quantum gravity.
The question whether or not Dirac and reduced phase space quantization are equivalent remains
open \cite{Ashtekar:1982wv, Carlip:1993zi, Vanrietvelde:2018pgb, Hoehn:2018whn}.

In the setting of AdS/CFT duality, the usual presumption is that the algebra of boundary observables 
is {\it identical} to the algebra of CFT observables. 
Moreover, the completeness of the boundary algebramight be a generic feature of quantum gravity \cite{Raju:2019qjq},
at least in a superselection sector \cite{Marolf:2012xe}.
If that is so, then, in a sense, the boundary information paradox is resolved simply by the fact that all information always resides at the boundary. 
However, our aim is to learn something about the Hilbert space and observables of quantum gravity, by confronting the the paradox from the bulk viewpoint:  the physical Hilbert space is, according to Dirac quantization, a subspace of a kinematic Hilbert space, which itself is, as in non-gravitational quantum field theory, roughly a tensor product of local factors. The boundary observables act on the boundary kinematic factor. To resolve the  BIP, then, one must understand how it is that certain ``bulk observables" can {\it also} correspond to boundary observables,\footnote{And how, perhaps, boundary observables could be complete.} without violating quantum mechanics.
To this end, it does not help to simply invoke completeness of the CFT observables. 

The rest of the paper is structured as follows. In Sec.~\ref{BU} the argument for BU is
recalled, and we identify analyticity of time evolution as a key distinction between the 
quantum and classical cases. Sec.~\ref{BIP} poses the BIP, and Sec.~\ref{Resolving} posits our proposed resolution of the paradox, which involves the notion of the ``nebulon," which is our name for the
nebulous operators of the boundary algebra that redundantly encode information about physics in the bulk. 
The purpose of Sec.~\ref{nebulon} is to explore more specific descriptions of the nebulon, both nonperturbative and perturbative, as well as to answer an objection from Ref.~\cite{Almheiri:2013hfa} to the
proposal that ``nebulous bulk degrees of freedom"  could resolve the black hole information paradox.
In Sec.~\ref{realities} we explain how it could be that the
nebulon avatar mechanism is not precluded by the (algebraic) no-cloning theorem.
In Sec.~\ref{infaller} we recall the infaller paradox, which is special to black holes, and 
we argue that as stated heretofore it is ill-posed in quantum gravity, because it ignores
essential aspects of the nature of the Hilbert space and observables in quantum gravity. 
Finally, we close in Sec.~\ref{remarks} with a few remarks.

\section{Boundary Unitarity}
\label{BU}

The Hamiltonian of a classical theory with diffeomorphism symmetry, such as general relativity, is a linear combination of constraints, plus a boundary term if the spatial manifold $\Sigma$ has a boundary 
$\partial\Sigma$ \cite{Regge:1974zd},
\begin{equation}
    H = \int_{\Sigma} N_{\mu} C^{\mu} + H_{\partial}\,.
\end{equation}
Here $N_{\mu}$ are the lapse and shift functions, $C^{\mu}$ are the constraints, and $H_{\partial}$ is the boundary term. If $N_\mu$ approaches an asymptotic time translation at the boundary, $H$ generates that time translation 
via Hamilton's equations. 
The constraints arise because of diffeomorphism symmetry. The Poisson brackets of the constraints close on the constraints (the constraints are ``first class"), 
hence the constraints generate ``gauge" transformations corresponding to diffeomorphisms. 
Any gauge-invariant observable $\mathcal{O}$ Poisson-commutes with the constraints $\{ \mathcal{O}, C^{\mu} \} = 0$
(up to terms that vanish with the constraints),
so the evolution of observables with respect to asymptotic time-translations 
is governed entirely by the boundary term $H_{\partial}$:
\begin{equation}\label{Odotclassical}
    \frac{d\mathcal{O}}{dt} = \{\mathcal{O}, H\} = \{ \mathcal{O} , H_{\partial}\}.
\end{equation}
So far this is all classical physics. 

In the quantized version of the foregoing structure, the canonical variables become operators, the Poisson bracket relations become commutator relations, and {\it presumably}, if the diffeomorphism symmetry survives quantization, the algebra of the constraints still closes on the constraints. Of course it is well known that
the quantization of general relativity cannot by itself make sense, without some ``UV completion" of the theory. 
But if the diffeomorphism symmetry survives this UV completion it 
is plausible
that the Hamiltonian is still a boundary term, 
and that gauge invariant observables evolve by commutation with this Hamiltonian, according to the equation
\beq\label{Odot}
  \frac{d\mathcal{O}}{dt} =\frac{1}{i\hbar} [ \mathcal{O} , H_{\partial}],
\eeq
using the Heisenberg picture. In Sec. \ref{Resolving} we discuss the possibility that in the 
UV completion of the gravity theory the Hamiltonian is no longer localized at the boundary, and we argue that even if so, that in itself likely does not eliminate the BIP.

\subsection{Boundary observables and unitarity}

Boundary unitarity refers to the time evolution of boundary observables, by which we mean (gauge invariant, self-adjoint) observables that can be built from the fields in the intersection of any neighborhood of a bulk Cauchy slice with any neighborhood of the boundary. 
We refer to the algebra generated by these as the
{\it boundary algebra} \cA~associated with the given (asymptotic) Cauchy slice. 
The Hamiltonian---at least at the level of EFT---is in the boundary algebra, as is the commutator of the Hamiltonian with anything in the algebra. 
The time derivative \eqref{Odot} is 
therefore in \cA, which leads to the expectation that \cA~evolves unitarily into itself. 
This property was termed ``boundary unitarity'' (BU) in \cite{Marolf:2008mf}. 
BU implies that the information contained in boundary observables is time independent. 

Note that we are {\it not} assuming that the boundary algebra includes all
observables, although that might be the case.\footnote{It was recently argued in Ref.~\cite{Raju:2019qjq}
that in canonical quantum gravity, if a Reeh-Schlieder property holds for boundary operators, and if
the projection on the vacuum lies in \cA, then indeed \cA~ would coincide with the full algebra of observables. Also, in the setting of AdS/CFT duality, it is tempting to identify the CFT observables,
which are complete for the theory, 
with the boundary observables.}
Moreover, if there is more than one boundary of the spacetime, 
then there is more than one boundary algebra, so the observables at one boundary 
are clearly not complete,
but this does not change anything essential in our considerations.
There would be one term in the Hamiltonian for each boundary, 
and the algebra of observables at any one boundary would evolve into itself. 

But before getting carried away with the implications of BU and the paradox it raises, 
we must examine more closely the justification for the expectation that \cA~evolves unitarily. 
Something more than the argument already given is clearly required, because the 
corresponding {\it classical} statement is {\it not} true \cite{Marolf:2008mf,Waldprivate}. There are 
solutions of the Einstein equation that agree {\it exactly}, for example, with the Schwarzschild-AdS solution
outside some radius on some time slice, and which nevertheless have disturbances which propagate out to the boundary at a later time   \cite{Isenberg:2005xp} (see also \cite{2000CMaPh.214..137C,Chrusciel:2003sr,Corvino:2003sp}). 
The arrival of such disturbances would clearly constitute new information in the boundary observables, so would violate the classical version of BU. 
In fact, the quantum case 
has a key property that the classical case lacks: analyticity of time evolution.

\subsection{Classical case}
The classical evolution equation \eqref{Odotclassical} is closely analogous to the quantum one
\eqref{Odot}, and the
reasoning we applied above for the latter applies to the former: the rate of change of an observable is another element of the boundary algebra. However, as just noted, that must not in general be sufficient to infer that this property is shared by the time-dependent solution to this equation. That solution can be written formally, 
as 
\beq\label{solclassical}
{\cal O}(t)=e^{t\{\cdot,H\}}{\cal O}(0),
\eeq
where $\{\cdot,H\}$ denotes the operation of taking the Poisson bracket with the function to its right, placed in the first entry of the bracket. Expanding the exponential, \eqref{solclassical} becomes an infinite series of nested Poisson brackets. The question, then, is whether this series converges, for all $t$, to the actual solution to \eqref{Odotclassical}, on each integral curve of the Hamiltonian vector field on phase space. The answer is evidently no, not in general: not all solutions are analytic in $t$ with infinite radius of convergence, so in general the series does not converge to the solution for all $t$. Thus, although it {\it formally} appears from \eqref{Odotclassical} that the values of the boundary observables at one time determines the values at any other time, that fails to be the case.

Let us illustrate this concretely using a massless scalar field $\phi(x,t)$ in $1+1$ spacetime dimensions.
The Hamiltonian is $H = \int dx[\half\pi^2 + \half (\partial_x\phi)^2]$, the conjugate momentum is 
$\pi=\partial_t\phi$, and for the observable we take ${\cal O}=\phi(x_o)$, for some fixed $x_o$.  
Consider now a right-moving solution: $\phi(x,t) = f(x-t)$, for some function $f$. This solution
is determined by the initial data $\phi(x,0)=f(x)$ and $\pi(x,0) = -\partial_xf(x)$. For $n$ nested commutators,
$\{\{\{\phi(x),H\},\}\dots,H\}\phi(x)$ is equal to 
$\partial_x^{n-1}\pi(x)$ for odd $n$, and $\partial_x^{n}\phi(x)$ for even $n$.
Using the initial condition we thus find, on this particular trajectory,
\beq
e^{t\{\cdot,H\}}\phi(x_o,0)= \sum_{n=0}^\infty \frac{1}{n!}(-t)^n\partial_x^n f(x_o), 
\eeq
which is the Taylor series for $f(x_o-t)$.  If $f(x)$ is not analytic with infinite radius of convergence, then
for some $t$ the series will fail to agree with $\phi(x_o,t)$.
 For example, suppose $f(x)$ is a
bump function, $\exp[-1/(x^2-1)]$ for $|x|<1$ and $0$ for $|x|\ge1$, and choose $x_o=2$. Then the series 
converges to zero for all $t$, whereas $\phi(2,t)$ is nonzero for $1<t<3$. Alternatively, if we choose $x_o=0$, then the
series converges to the correct solution  only for $|t|<1$. 

\subsection{Quantum case}

The classical analysis suggests that, if BU is to hold, then there must be something about quantum mechanics that enforces analyticity,
or something like it, for the time dependence of the boundary observables. This is indeed the case. 
The solution to \eqref{Odot} takes the form
\beq\label{Ot}
{\cal O}(t)=  e^{itH/\hbar}{\cal O}(0)e^{-itH/\hbar}.
\eeq
We assume that $H$ is self-adjoint, so that 
the exponential $e^{-itH/\hbar}$ is well-defined in terms of the spectral projections of $H$ \cite{reed1981functional}.
Moreover, the exponential is a bounded operator. To ensure that the operator on the right hand side of \eqref{Ot} is defined everywhere, 
we assume that also ${\cal O}$ is bounded. Then \eqref{Ot} is indeed the solution to \eqref{Odot} wherever the latter is defined.\footnote{And \eqref{Ot} replaces \eqref{Odot} wherever the rhs of \eqref{Odot} is {\it not} well defined, as is usual in quantum mechanics.} 
And, since it is a product of three operators each of which lies in \cA, it too lies in \cA. 

While the preceding argument for BU is logically sufficient, we can give a more constructive account by
considering expectation values $\la\psi|{\cal O}(t)|\psi\ra$,
and restricting to energy-bounded observables. This will also exhibit clearly the difference between the classical and quantum cases. 

By inserting complete sets of $H$ eigenstates on either side of ${\cal O}(0)$, we obtain 
using \eqref{Ot} an expression for the expectation value as a double sum,
\beq\label{<O>}
\la\psi|{\cal O}(t)|\psi\ra=\sum_{m,n}e^{i(\o_m-\o_n)t}{\cal O}_{mn}\la\psi|m\ra\la n|\psi\ra,
\eeq
where $H|n\ra=\hbar \o_n|n\ra$ and ${\cal O}_{mn}=\la m|{\cal O}(0)|n\ra$. If the spectrum of $H$ includes a continuum, then the summation over the eigenstates includes integration. 
Since the exponentials are analytic functions of $t$, the expectation value $\la\psi|{\cal O}(t)|\psi\ra$ is a sum of 
analytic functions. Were there a finite number of terms in the sum, it too would be analytic, but there are in general an infinite
number of terms, so we must be more careful. 

If, as we shall assume, the Hamiltonian is bounded below, the range of the frequencies summed over in \eqref{<O>} is bounded below.
To ensure analyticity, we can cut off the energy at some upper bound $\Lambda$, and consider the
energy-projected observable ${\cal O}_\L:=P_\L{\cal O}P_\L$. Since the Hamiltonian is in \cA, so is $P_\L$,
because $P_\L$ is a function of the Hamiltonian,
\begin{equation}\label{PL}
    P_{\L} = \Theta{(\L-H)},
\end{equation}
where $\Theta$ is the Heaviside step function and $\L$ is the energy cutoff.
Hence the projected observable ${\cal O}_\L$ is also in \cA,  provided that ${\cal O}$ is. 
Moreover, the expectation value $\la\psi|{\cal O}_\L(t)|\psi\ra$ is analytic in $t$. To see this, just note that it is the diagonal value of a
double Fourier transform, $F(t,t')=\int d\o\int\d\o' e^{-i\o t}e^{-i\o't'}f(\o,\o')$, where $f(\o,\o')$ has compact support in each argument. Such a function is analytic in each argument, and hence the function $G(t):=F(t,t)$ is analytic in $t$. 

For each energy cutoff $\L$ we thus have an algebra of boundary observables ${\cal A}_{\partial}^{\L}$ whose expectation values in any state evolve  analytically. It follows that the collection of all time derivatives of the expectation values at one time determines the values at any other time. This is a concrete formulation of boundary unitarity.  The statement holds for any cutoff $\L$, hence for each $\L$ there is a BU property.
Moreover, we can take $\L$ arbitrarily large, and so come arbitrarily close to encompassing all boundary observable values for finite energy states.
{\it It is the unitarity of time evolution in quantum mechanics that leads to the analyticity property 
noted here.} In contrast, the classical argument for boundary unitarity fails precisely on account of the lack of analyticity in time. 

It is worth noting 
that the fact that the energy-projected observable 
$P_\L \mathcal{O} P_\L$ is in the boundary algebra is possible only in the presence of gravity.  
In local, relativistic field theory without gravity, by contrast, 
an energy-projected version $P_\L A P_\L$
of a local observable $A$ is {\it never}  localized to any open proper subregion $U$ of spacetime.
To see this, suppose $P_{\L} A P_{\L}$ were localized to some 
region $U$, and choose any observable $B$ 
localized in the causal complement $U^c$ of $U$. 
Microcausality then implies
\begin{equation}\label{PAPB}
    (P_{\L} A P_{\L}) B |\psi\rangle = \pm B (P_{\L} A P_{\L}) |\psi\rangle,
\end{equation}
where $|\psi\ra$ is any state, and 
the minus sign applies when both of the operators are fermionic.
Equation (\ref{PAPB}) implies in particular that, for all $B$ localized in $U^c$, 
the state $B (P_{\L} A P_{\L}) |\psi\rangle$ 
has energy content bounded by $\L$. This contradicts the 
generalized Reeh-Schlieder theorem which
asserts that, by acting on a state of finite energy content with operators localized in $U^c$, 
we can approximate {\it any} other state arbitrarily well \cite{Summers:2008yd,Witten:2018lha},
including ones with energy content greater than $\Lambda$. 

\section{Boundary Information Paradox}
\label{BIP}


The fact that bulk particles can propagate to the boundary raises a puzzle for 
BU \cite{JACOBSON:2013ewa}: how can the boundary algebra evolve unitarily into itself, when ``new information" can arrive at the boundary?  
When such a particle arrives at the boundary and influences the boundary observables, 
it appears that the 
boundary algebra is not evolving autonomously. 
 We shall call this puzzle the {\it boundary information paradox} (BIP). 
 An example of it arises when a black hole emits Hawking radiation which is correlated with the field behind the horizon, yet which reaches the boundary. However, the BIP arises even when no black hole is present. 
If a bulk particle 
propagates to the boundary after having been dynamically 
correlated with bulk degrees of freedom,
this puzzle is like the Hawking radiation one. The entropy of the boundary algebra seems to increase.
Moreover, there is even a puzzle without such correlation, 
because the arrival of a bulk particle at the boundary seems to introduce new information there.
The BIP focuses attention on an aspect of the black hole information paradox that deserves more attention than it has received, namely, the puzzle of {\it continuous unitary evolution of boundary observables}.
 
Consider, for example, a scattering type 
process in which particles are injected from the boundary
and interact in the bulk, 
forming a state that emits a massless particle while leaving behind a long 
lived resonance that is correlated to the early emitted particle. 
The arrival of the decay particle at the boundary can be detected by the measurement of some boundary observable $\mathcal{O}$, which is then correlated with the remaining resonance. This appearance of correlation between a boundary observable and a non-boundary one violates the fact that the boundary algebra evolves unitarily into itself. 

This scenario is rather similar to one in which the long lived resonance is a black hole. The key difference is that, in the case of the black hole, the early Hawking radiation is correlated not just to the remaining resonance, a.k.a. the black hole, but specifically to partner field modes behind the horizon of the black hole.
Also, the infinite redshift at the horizon implicates a mixing of bulk 
UV and IR degrees of freedom in the process. 
However, from the viewpoint of boundary unitarity, these distinctions are immaterial.  

We speak of dynamical correlation, and avoid the 
concept of ``entanglement," because the latter
refers to a state in a Hilbert space composed of two or more factors. In the setting of quantum gravity, however, a tensor factorization according to spatial localization is not available \cite{Donnelly:2015hta, Donnelly:2016rvo, Giddings:2018umg, Casini:2013rba}. Moreover, the particles or field quanta referred to in the previous paragraph are not gauge invariant with respect to diffeomorphisms; and, if they are gravitationally dressed in order to become so, then they are no longer spatially localized. 
But we have framed the BIP so as not to presume spatial factorization of the Hilbert space or any 
particular gravitational dressing.
The algebra of boundary observables is defined without invoking
a boundary factor of the quantum gravity Hilbert space.

Despite the absence of a factorized Hilbert space, 
given an algebra of quantum observables one can define 
the {\it algebraic entropy} of a global quantum state restricted to the subalgebra \cite{ohya2004quantum, Harlow:2016vwg}, 
which generalizes the notion of the von Neumann entropy of a state restricted to a tensor factor.  
The BIP can be phrased in terms of the behavior of this algebraic entropy:
If the Hamiltonian is an element of the subalgebra, then the algebraic entropy of the subalgebra cannot change in time. BU thus implies that the entropy of the boundary algebra never changes, and yet 
the arrival of bulk particles at the boundary appears to entail entropy change.

\section{Resolving the BIP}
\label{Resolving}

We can see two possible routes to resolving the BIP:
\ben
\item The Hamiltonian is {\it not} in \cA.
\item \cA~contains an avatar of any bulk degree of freedom with which it can be correlated.
\een
The argument for BU, reviewed in Section \ref{BU}, is that, 
on account of diffeomorphism invariance, the 
Hamiltonian {\it must} be in \cA. However, 
the boundary algebra was defined using a notion of locality that makes sense within field theory but may cease to be meaningful in the UV completion of the gravitational theory. For instance, strict localization of the Hamiltonian to any neighborhood of the boundary may not be possible in string theory, because of the extended nature of strings. In this case, the Hamiltonian would couple EFT boundary observables to non-boundary observables, so observables initially in \cA~ would evolve to observables not contained in \cA, and the BU argument as it stands would fail.\footnote{In the AdS/CFT context, 
it is plausible that there is no nontrivial subalgebra of the CFT that is invariant under time evolution. (We thank Hong Liu for a discussion on this point.) 
If this is the case, then either \cA~must be equivalent to the entire CFT algebra, or the Hamiltonian must not lie in \cA.} 

That said,  
this failure would not necessarily provide a resolution to the BIP, because the observables involved in the BIP may be restricted to the low energy effective field theory sector, and it is not at all clear that a tiny amount of dynamical mixing with observables outside of \cA~ would suffice to account for the non-conservation of information in 
\cA. 
Note also that the asymptotic redshift appears to help suppress such mixing. A finite Killing energy, 
defined with respect an asymptotic global time translation Killing vector $\partial_t$, 
corresponds to a vanishing {\it proper} energy in the frame defined by the unit vector $\sim r^{-1}\partial_t$, 
where $r$ is the asymptotic area radius coordinate. Hence there is not enough energy to create
``stringy" excitations near the boundary.

For the remainder of this paper, we will focus on the second route, namely, the possibility that 
\cA~ already contains an avatar--- i.e.\ an image, copy, or representative---of any 
component of the bulk degrees of freedom
needed to preserve the information available at the boundary. 
Related approaches to solving the black hole information paradox have been discussed before,
for example in the form of black hole complementarity and ``$A = R_B$" \cite{Susskind:1993if, Susskind:1993mu, Lowe:1995ac, tHooft:1984kcu}.
These have been critiqued  as non-viable, generally for reasons related to the ``no-cloning theorem"
of quantum mechanics, or contradictions resulting from the presence of the same quantum information in more than one location. It seems to us that, if formulated carefully, no contradictions or violations of no
cloning need arise. In the next section we articulate a little more explicitly the nature
of our proposed resolution, and in the following section we confront the no-cloning issue.

\section{The Nebulon}
\label{nebulon}

The notion that such a an avatar must be available at the boundary to purify Hawking radiation was 
proposed in \cite{JACOBSON:2013ewa}, in which the continuous unitarity feature of the BIP was emphasized.
This proposal was criticized in \cite{Almheiri:2013hfa}, which argued that 
``nebulous bulk degrees of freedom" could play no role in evading a version of
the black hole information paradox involving an auxiliary system that extracts and converts 
Hawking quanta into excitations of a separate system attached to spacetime. (That criticism will
be addressed below.) 
Inspired by that denigration, we are moved to embrace the term, and here refer to the required avatar as the {\it nebulon}. 

\subsection{Who is the nebulon?}

Since it is diff invariance that leads to the BIP
in the first place, we expect that diff invariance 
should lie at the root its resolution. The impact of diff invariance on the structure of the
quantum gravity state space is indeed very strong. 
Classically, the Hamiltonian constraint builds the nonlocal aspects of gravity,
like the link between sources and geometry in the bulk and 
multipole moments measurable at infinity,
into the structure of the phase space. 
At the quantum level, in the setting of Dirac quantization, the Wheeler-DeWitt equation
presumably does the same job, and more: it encodes also any evolution with respect 
to internal clocks. So a likely candidate for the nebulon mechanism is the Wheeler-DeWitt equation.

Another key aspect of the nebulon may be simply the time evolution of the boundary observables.
For example, a bulk particle arriving at the boundary appears to 
bring new information.  But, if (thanks to diff invariance) BU indeed holds, 
that information must have already been available at the boundary
before the arrival of the particle. Long ago, in the setting of AdS/CFT, 
the boundary state that anticipates the arrival of a particle was referred to as a {\it precursor} \cite{Polchinski:1999yd, Susskind:1999ey, Freivogel:2002ex},  the 
precursor being a CFT state. It is easy, in a sense, to describe the precursor observable: we can ``simply" time evolve the boundary particle detection observable backwards in time, to express it as a boundary observable at any earlier time. At the earlier time, it is presumably a highly ``scrambled" looking observable, which would be difficult to recognize as the precursor of a simple particle detection, and yet BU guarantees its existence in \cA. 

\subsection{Perturbative nebulon?}\label{Sec:PapaRaju}

The nebulon has already appeared, and not so nebulously, in the context of AdS/CFT,
at the level of effective field theory, at zeroth order in Newton's constant, 
in the guise of expressions for bulk operators in terms of boundary operators.
We have in mind for example the well known HKLL  \cite{Hamilton:2006az} reconstruction of a bulk scalar field, which uses the scalar field equation of motion
to express a bulk field in terms of its boundary values in the region of the boundary that
is spacelike related to the bulk point. Another example is
the less well known BBPR  reconstruction  \cite{Banerjee:2016mhh, Raju:2018zpn}, which 
uses the Reeh-Schlieder property \cite{Summers:2008yd} 
that arises from the strong vacuum entanglement, together with 
the fact that the Hamiltonian---and therefore the projection onto the ground state---is
in the boundary algebra. 
Although both of those examples were studied in the context of AdS/CFT, and mostly with pure AdS as a background, they do not, as far as we can see,
require anything beyond the general setting of asymptotically AdS quantum gravity.

As mentioned, these ``bulk reconstruction" approaches work, initially, without accounting for 
gravitational coupling, and without enforcing diffeomorphism invariance. 
Perturbative corrections at lowest order in $O(1/N)$, where $N$ is the central charge of the CFT
presumed dual to the quantum gravity theory, were constructed 
by Kabat and Lifshitz (in a Fefferman-Graham gauge) 
in \cite{Kabat:2013wga}.
To the extent that any quantum gravity theory is dual to a CFT, this can presumably be 
viewed as a gravitational perturbation expansion in $(L_{\rm Planck}/L_{\rm AdS})^{D-2}$
where $D$ is the bulk spacetime dimension.   

In another approach \cite{Marolf:2008mf}, Marolf argued (without invoking AdS/CFT)
that, using the bulk equations of motion, bulk operators can be expressed in terms of boundary 
operators lying to the past of the bulk point.
Moreover he argued that 
the dependence on the boundary algebra can be squashed down to the infinitesimal
neighborhood of a single Cauchy slice,
using the fact that the Hamiltonian, and therefore the 
time evolution operator,  lies in the boundary algebra.
He inferred from this that,   
to all orders in perturbation theory, all bulk operators are represented in 
the boundary algebra at any time.\footnote{He allowed, however, that beyond perturbation theory it might 
not be the case that the boundary algebra is complete in this sense. But such incompleteness,
were it to be the case, would not be incompatible with BU, which refers only to the unitary evolution of the boundary algebra, be it compete or not.} 
We view this line of reasoning (which can also be applied to the HKLL reconstruction), 
as well as the BBPR  reconstruction  \cite{Banerjee:2016mhh, Raju:2018zpn}, 
as strongly supporting the nebulon hypothesis. 
However,  since the boundary Hamiltonian 
(when expressed in terms of the canonically normalized metric perturbation) 
is inversely proportional to the gravitational coupling $\kappa$, 
the operation of squashing to a single time slice, 
and the projection onto the ground state, are non-perturbative in the coupling,
so that the realization of the nebulon on a single time slice may be essentially nonperturbative.

It should be further noted that the perturbative analysis of ``gravitational splitting" 
in \cite{Donnelly:2018nbv,Giddings:2019hjc} casts doubt on 
the existence of a perturbative nebulon.\footnote{We thank Steve Giddings for 
calling our attention to this issue.}
 It was shown there that, 
to first order in $\kappa$, inequivalent states 
can be indistinguishable outside some compact set $U$.
These states are constructed in  \cite{Donnelly:2018nbv}
by acting on a ``split vacuum" 
with operators that are dressed so as to be gauge invariant to first order
in $\kappa$, and in such a way that observables localized 
outside of $U$ are sensitive to only the Poincar\'{e} charges of these
states. This seems to imply that the algebra of operators outside
$U$, and in particular the boundary algebra, does not contain observables
that can distinguish these different states, and hence that the boundary algebra
must {\it not} be complete. 
This conclusion is in tension with 
BU, and with what is known about bulk reconstruction in AdS/CFT,
but there is is no contradiction:  
the splitting analysis was perturbative on one time slice, 
whereas  the  boundary time evolution that plays a role in BU 
and bulk reconstruction, or translation of bulk operators to the boundary \cite{Giddings:2018umg}, 
is nonperturbative.

A caveat should be attached to any perturbative discussion of this sort:
the notion of a bulk operator satisfying bulk microcausality, i.e.\ commuting with operators at
spacelike related points, apparently cannot be defined on the entire Hilbert space 
of the theory  \cite{Almheiri:2014lwa,Harlow:2016vwg}. 
Instead, it was argued in  \cite{Almheiri:2014lwa}
that local bulk observables are defined only on a semiclassical subspace of the 
full quantum gravity theory Hilbert space.
This subspace, was called the \textit{code subspace}, because 
it appears that the encoding of bulk quantum information in the boundary 
algebra is redundant, in the manner of a quantum error-correcting code (QEC). 
In particular, the HKLL construction can be carried out restricting to a single ``causal wedge"
of AdS (and this can be generalized to a reconstruction in an entanglement wedge of 
a deformed spacetime \cite{Dong:2016eik}). Since a given bulk point can lie in many, overlapping causal 
wedges, one obtains many inequivalent boundary representatives for the same bulk operator. 
Be that as it may, this redundant encoding  aspect of holography presents no
problem for BU. 

\subsection{AUX vs. Nebulon}

Ref.~\cite{Almheiri:2013hfa}, `` An Apologia for Firewalls," gave an argument involving an auxiliary system, AUX, purporting to demonstrate that purification of Hawking radiation by nebulous degrees of freedom could not solve the BIP, since it 
would be inconsistent with unitary evolution. The  AUX system interacts with asymptotically AdS spacetime at the boundary, coupling to it only via the boundary value of a bulk scalar field, i.e.\ AUX couples only to a single trace, scalar primary operator in the CFT dual to the gravity theory. Via this interaction, a scalar Hawking quantum can be absorbed, its energy being transferred to an infinitely large AUX reservoir from which it will never return. If then a large black hole is formed from a pure state, it can evaporate completely,
depositing all of the initial energy in the AUX reservoir. Since the entire evolution is unitary, the entropy of the AUX state must vanish.\footnote{In the scenario of Ref.~\cite{Almheiri:2013hfa} the black hole
was only allowed to evaporate down to another stable black hole, much smaller than the initial one. For rhetorical simplicity, here we allow the black hole to evaporate completely.} 

The argument in \cite{Almheiri:2013hfa} was that since AUX couples only to the scalar Hawking quanta, and not to any nebulous degrees of freedom that might have purified the Hawking quanta, the AUX quanta must inherit the entropy of the Hawking quanta, and therefore AUX can not wind up in a pure state, unless the Hawking quanta taken together, by themselves, are in a pure state.
The Apologia draws the conclusion that there must be a firewall at the black hole horizon that precludes the entanglement of Hawking quanta with behind-horizon partners.
 However, this conclusion does not follow from the assumptions. As the black hole shrinks, and the temperature decreases, the scalar field quanta interact with the gravitational field, and hence, if only indirectly, with everything. Any entanglement between scalar and nebulon degrees of freedom will be channeled into the extracted scalar quanta as the system cools. That is,
 the entanglement between the scalar quanta and the nebulons will be extracted into AUX through the scalar channel.
 A simple analogy makes this 
transparent:  Consider a gas mixture of helium and neon in a zero entropy state that looks like a high temperature equilibrium state, and imagine cooling the gas by coupling an auxiliary refrigerant AUX, able to absorb kinetic energy of only the neon atoms at the walls of the container. Despite the entanglement of the neon with the helium atoms, and despite the fact that AUX couples only to the neon, it is obvious that once the atoms have cooled to zero temperature (and assuming the ground state has
zero entropy), AUX winds up in a pure state. 

\section{No-cloning constraints}
\label{realities}

The nebulon avatar 
mechanism for resolving the BIP posits that the same quantum information 
can ``simultaneously" 
take more than one form in a single global quantum state. For example, a particle can be 
correlated with two different subsystems, using either of which the same correlation 
information can be measured. The idea behind this is that the Wheeler-DeWitt equation 
may enforce redundant encoding of the quantum information. 
This mechanism was proposed in \cite{JACOBSON:2013ewa}, and illustrated with an example 
involving four spins. However, that example involves only 
abelian observable algebras, which are inadequate for quantum state tomography
(and for describing the physics at hand),
and linear algebra imposes very strong constraints on the structure of redundant encoding for 
nonabelian algebras. Such constraints are captured by 
the {\it algebraic no-cloning theorem} (ANC) \cite{Almheiri:2014lwa}.
In this section we review the ANC, and then explain how the nebulon avatar mechanism 
might evade it.

\subsection{The algebraic no-cloning theorem}

In the framework of
Dirac constrained quantization, the 
physical Hilbert space $\mathcal{H}_{phys}$
is the subspace of a ``kinematical'' Hilbert space $\mathcal{H}_{kin}$ 
that is annihilated by the diffeomorphism constraints,
\begin{equation}
    \mathcal{H}_{phys} = \bigg \{ |\psi\rangle \in \mathcal{H}_{kin} \bigg| C |\psi\rangle = 0 \bigg \},
\end{equation}
where $C$ represents the constraints, which generate gauge transformations.
Now suppose  $\mathcal{H}_{kin}$ is composed of two tensor factors:
\begin{equation}
    \mathcal{H}_{kin} = \mathcal{H}_{A} \otimes \mathcal{H}_{B}.
\end{equation}
In the gravity setting, we have in mind that one factor corresponds to the kinematic degrees of freedom in the 
near boundary region, and the other corresponds to the 
degrees of freedom deeper in the bulk. But at the moment we are just concerned with the algebraic structure.

We are interested in an algebra $\mathcal{A}$ of physical operators, i.e.\ 
operators that commute with the constraints $C$ and therefore map
the space of physical states to itself. 
The ANC states: 
\begin{quote}
\it If $\mathcal{A}$ is measurable on {\it both} tensor factors,   
then it can only be an abelian algebra.
\end{quote}
What it means for $\mathcal{A}$ to be measurable on the tensor factor  
$\mathcal{H}_{A}$ is that, for every operator $\mathcal{O} \in \mathcal{A}$, 
there exists an operator $\mathcal{O}_{A}$ acting on $\mathcal{H}_{A}$ such that
\begin{equation}\label{rep}
    \mathcal{O}|\psi\rangle = (\mathcal{O}_{A} \otimes I_{B}) |\psi\rangle
\end{equation}
for all physical states $|\psi\ra\in \mathcal{H}_{phys}$.\footnote{This is equivalent to 
the requirement that the expectation values  of 
$\mathcal{O}$ and $\mathcal{O}_{A} \otimes I_{B}$ agree for
all physical states.} 
A similar definition applies to measurability in the other tensor factor, $\mathcal{H}_{B}$.
The proof of the ANC is trivial
(as is that of the classic no-cloning theorem of quantum mechanics \cite{Wootters:1982zz}):
Consider any two operators in the algebra, and  
represent one of them on one tensor factor as in \eqref{rep}, 
and the other one on the other tensor factor. 
The two representatives commute, simply because they act on different tensor factors. By
hypothesis, the action of the 
representatives on the physical states is the same as that of the 
original operators, so this implies that all of the operators in $\mathcal{A}$ must commute when 
acting on $\mathcal{H}_{phys}$.

\subsection{How gravity bypasses the ANC}
The ANC appears to present an insurmountable obstruction to our proposed resolution of 
However, the situation in quantum gravity does not conform to all of the assumptions of the
the ANC theorem. To identify the wrong assumption, we consider the 
algebra of Poincar\'{e} charges (ADM energy, momentum etc.) of 
asymptotically flat spacetime.
This algebra is nonabelian, but every operator in it 
can be expressed both as a bulk integral and as a surface integral. It would thus seem that the Poincar\'e charges are measurable in both the near-boundary tensor factor and the bulk tensor factor. Indeed, this fact was recently exploited by Donnelly \cite{Donnelly:2018qya}, in the linearized theory, 
to implement quantum state tomography of single-particle bulk states via boundary measurements.

How is the above observation  about the Poincar\'{e} charges to be reconciled with the ANC? We suppose 
that the reconciliation must be that the ``second copy" of the observables, i.e.\ the bulk version, is not actually fully contained on a separate (kinematic) tensor factor, 
because the bulk integral must actually go all the way to infinity,
since otherwise it is not even gauge invariant. Similarly, in the case of the nebulon, 
we suppose it is because of gravitational dressing required to turn the resonance into a gauge-invariant observable. It has been proved by Donnelly and Giddings \cite{Donnelly:2016rvo}, using a perturbation expansion in Newton's constant,\footnote{The expansion is really in the ratio of a relevant energy to the Planck energy. As such, the result applies to the observables acting on a class of low energy states.}
that 
for any operator with nonzero Poincar\'e charge
such gravitational dressing 
always extends to infinity.\footnote{Poincar\'{e} charges are given by surface integrals at infinity, so that if an operator does not extend to infinity, it will commute with all the Poincar\'e charges and hence, since the Poncar\'e group has no abelian invariant subgroup, the operator cannot carry any Poincar\'e charge.}
Put slightly differently, this dressing theorem---which has also been extended to asymptotically
AdS gravity \cite{Giddings:2018umg}---implies that the division of kinematic Hilbert space into a near-boundary tensor factor and bulk tensor factor does not pass to the level of the algebra of gauge invariant observables. The theorem asserts that there do not exist any gauge-invariant observables (with nonzero asymptotic charges) acting on the bulk tensor factor alone, since all such observables extend to the boundary.

\section{Infaller paradox}
\label{infaller}

We turn in this section to a brief consideration and critique of the infaller paradox \cite{Bousso:2012as,Almheiri:2013hfa}.
The infaller paradox  is based on the 
supposition that black hole evaporation is a unitary process, and the argument goes as follows. 
After a black hole has radiated more than half of 
the Hawking radiation (i.e.\ past the {\it Page time}), 
each subsequent Hawking quantum $q$ must be essentially maximally
entangled with the earlier radiation.  Moreover, 
Alice, who is hanging around outside the black hole, can distill from the earlier radiation a special qubit $s$ 
that has extremely high probability of being maximally entangled with a particular subsequent Hawking quantum $q$. 
With this special qubit $s$ in hand, Alice can 
choose to measure either the joint state of $sq$ or the state of $\tilde q q$, where $\tilde q$ is the partner behind the horizon with
which $q$ is strongly entangled. This measurement takes place in a semiclassical region of spacetime, and in a setting
where gravitational effects are negligible, and quantum gravity subtleties can be ignored, so ordinary local quantum field theory should correctly describe the results of Alice's measurements. 
Subtleties of diffeomorphism invariance, dressed operators, and the quantum gravity Hilbert space, seem to be irrelevant for this measurement part of the argument. 
But the double entanglement of $q$ with both $s$ and $\tilde q$ is impossible in quantum
mechanics. So, assuming Alice can indeed possess the special qubit $s$, 
and barring a breakdown in local, low energy effective field theory (such as that explored in \cite{Giddings:2017mym}), 
one must conclude that $q$ cannot after all be strongly entangled with $\tilde q$.
But such entanglement is inevitable in the regular vacuum, so it follows that 
there must be a ``firewall" rather than a smooth vacuum at the horizon.

The weak link in this infaller paradox reasoning is the argument that Alice can possess the special qubit $s$. 
This link goes far beyond the assumption that local quantum field theory should hold good in a local region 
of spacetime where Alice's measurements of $sq$ or $\tilde q q$ take place. 
Indeed, it relies on treating the Hilbert space of black hole plus radiation 
like an ordinary quantum system, without gravity. The Hilbert space is taken to have the form 
${\cal H}_{\rm bh} \otimes {\cal H}_{\rm rad}$, and the radiation degrees of freedom are further factorized into the early radiation 
an the late radiation, ${\cal H}_{\rm rad} = {\cal H}_{\rm rad, e} \otimes {\cal H}_{\rm rad, l}$. 
Invoking Page's theorem \cite{Page:1993df,Harlow:2014yka}, the argument asserts that, past the Page time
(and assuming a randomness in the structure of the state), each of the later Hawking quanta 
in ${\cal H}_{\rm rad, l}$
must be essentially maximally entangled with the earlier ones in ${\cal H}_{\rm rad, e}$.
Further, the argument supposes that Alice knows precisely the initial state that formed the black hole,
and knows precisely the Hamiltonian of the system, so that in fact she knows precisely {\it how} $q$ is 
entangled with ${\cal H}_{\rm rad, e}$, and therefore she can perform a unitary operation on the early radiation 
to distill a special qubit $s$ that she knows must be maximally entangled with $q$.

Two factors related to diffeomorphism invariance 
render this argument dubious in the gravitating case. First, as discussed above, the Hilbert space is not factorized in the way that has been assumed. This means not only that the quantum states cannot be characterized as
they are in the formulation of the paradox, but also the very assumption that the Hawking radiation must be self purifying,
with no involvement of nebulons, is not justified.
Second, the protocol  for distillation on the early radiation must be diff invariant; in particular, the identification of the 
spacetime location of quanta acted on, and how they are acted on, must be  diff invariant. But diff invariant observables (with nonzero asymptotic charges) require dressing that reaches to the boundary \cite{Donnelly:2016rvo,Giddings:2018umg}.
This means that the distillation protocol implicates both local and global 
gravitational degrees of freedom other than the ideal QFT Hawking quanta to which the original argument applies,
and it calls into question whether Alice could possibly carry out the distillation, even in principle. 
Moreover, the Hawking quantum with which a particular special distilled qubit is supposed to be perfectly entangled must also be identified in a gauge invariant fashion via some gauge invariant observable.
Unlike for Alice's measurements themselves on the Hawking quanta, 
it is incorrect to argue that these quantum gravitational effects are irrelevant. 

In short, the effects of gravity and diffeorphism invariance turn the distillation protocol of Alice from a quasilocal to a global procedure, and although the effects that make it global may each be perturbatively small in the gravitational coupling, 
the protocol requires her to keep track of a huge amount of information, to very high precision. It seems perfectly possible that the global aspect of her task cannot be neglected, and that the upshot is that she simply cannot possess a special qubit that would violate local QFT near the horizon.
That is, the delicate task of distilling the special qubit $s$ and identifying the particular Hawking qubit $q$ with which it is supposed to be maximally entangled may be not only exceedingly challenging technologically, but may simply be impossible in principle, on account of the gravitational physics.

\section{Closing Remarks}
\label{remarks}

The main points we would like to emphasize are the following:
1) The boundary unitarity argument of Marolf appears sound, and the fact that it does not 
apply in the classical limit is understood to result from a lack of analyticity that has no quantum 
counterpart. 2)  
The black hole information paradox is just a special case of a more general boundary information paradox.
3) The resolution of the paradox(es) does not require a breakdown of known physics such as a 
firewall at the horizon \cite{Almheiri:2013hfa} 
or a breakdown of local quantum field theory as in non-violent nonlocality \cite{Giddings:2017mym}, but rather just a better accounting for the impact of diffeomorphism invariance on the global structure of the Hilbert space and observables in quantum gravity. 

The role of diffeomorphism invariance has been almost completely ignored in the voluminous literature on the information paradox. This is
rather ironic given that, as made clear by the boundary unitarity argument, it is precisely diffeomorphism invariance that leads to the paradox in the first place. 
We have argued that,
as in the case of other famous paradoxes of physics
(the twin paradox, the Gibbs paradox, the EPR paradox\dots), 
 the resolution is found {\it not} by forbidding the paradoxical scenario,
nor by modifying the theory that led to the paradox---rather it is found by more 
deeply learning the lessons of the theory as to what is {\it observable}, 
and what is a {\it meaningful statement about the world}.

\section*{Acknowledgements}
We are grateful to R. Bousso, S. Carlip, W. Donnelly, D. Marolf, S. Giddings, D. Harlow, H. Liu, S. Raju, R. Wald, A. Wall, and many other colleagues for discussions on these puzzling, thorny issues. This work was supported in part by NSF grant PHY-1708139.

\bibliography{bibliography.bib}

\end{document}